\documentclass[5p]{elsarticle}
\usepackage{graphicx}
\usepackage{amsmath}
\usepackage{amssymb}
\usepackage{enumerate}
\usepackage{url}

\begin{document}
\begin{frontmatter}
\title{A fast divide-and-conquer algorithm for indexing human genome sequences}
\author[a]{Woong-Kee Loh\corref{cor}}
\cortext[cor]{Corresponding author}
\author[b]{Yang-Sae Moon}
\author[c]{Wookey Lee}
\address[a]{Department of Multimedia, Sungkyul University}
\address[b]{Department of Computer Science, Kangwon National University}
\address[c]{Department of Industrial Engineering, Inha University}

\begin{abstract}
Since the release of human genome sequences, one of the most important research issues is about indexing the genome sequences, and the suffix tree is most widely adopted for that purpose. The traditional suffix tree construction algorithms have severe performance degradation due to the memory bottleneck problem. The recent disk-based algorithms also have limited performance improvement due to random disk accesses. Moreover, they do not fully utilize the recent CPUs with multiple cores.
In this paper, we propose a fast algorithm based on `divide-and-conquer' strategy for indexing the human genome sequences. Our algorithm almost eliminates random disk accesses by accessing the disk in the unit of contiguous chunks. In addition, our algorithm fully utilizes the multi-core CPUs by dividing the genome sequences into multiple partitions and then assigning each partition to a different core for parallel processing. Experimental results show that our algorithm outperforms the previous fastest DIGEST algorithm by up to 3.5 times.
\end{abstract}
\end{frontmatter}

\section{Introduction}
\label{sec:intro}

Due to recent advances in bio technology (BT), genome sequences of diverse organisms including human beings are collected into databases. The Human Genome Project (HGP), which had been initiated in 1990, released the human DNA sequences of approximately 3Gbp\footnote{bp stands for `base pair.' There are four bases, namely adenine~(A), cytosine~(C), guanine~(G), and thymine~(T).} size in 2003. Since the release, a lot of researches are under their way for harnessing the genome sequences. An essential research issue is about indexing large-scale genome sequences for efficient retrieving of genome subsequences of interest~\cite{alt97, bar08, gho09, hun02, pho07, pho08, tia05}. The suffix tree is most widely adopted for indexing genome sequences~\cite{bar08, bed04, che05, hun02, pho07, tia05}. In general, a suffix tree is created for a given string (or sequence) $X$ and enables efficient exact matching and approximate matching on substrings of $X$~\cite{gus97}. We explain the suffix tree in more detail in Section~\ref{sec:suffix}.

A lot of algorithms have been proposed for efficient construction of the suffix tree. Ukkonen's algorithm~\cite{ukk95} is the most famous one which, given a string of length $n$, constructs the corresponding suffix tree in $O(n)$ time. The algorithm implicitly assumes that $n$ is small enough so that the input string and the output suffix tree can be loaded in the main memory as a whole. However, genome sequences could be several million or billion times larger than the strings dealt with the traditional suffix tree construction algorithms such as Ukkonen's algorithm. Moreover, the suffix tree is about 10 $\sim$ 60 times larger than the input sequence~\cite{bar08, pho07, tia05}. Hence, the application of Ukkonen's algorithm for large-scale genome sequences should cause severe disk swap in and out, which is generally called {\it memory bottleneck problem\/} or {\it thrashing\/}~\cite{bar08, bed04, che05, hun02, pho07, tia05}. Actually, TOP-Q algorithm~\cite{bed04}, an extension of Ukkonen's algorithm, took seven hours for constructing the suffix tree for genome sequences of 40Mbp, which is much smaller than the human genome sequences, and it could not finish for genome sequences of 60Mbp~\cite{pho07}.

For coping with the memory bottleneck problem, a few disk-based algorithms have been proposed for constructing the suffix tree~\cite{bar08, che05, hun02, pho07, tia05}. Disks have much larger size than main memory at the lower cost; however, they require much longer access time up to several hundred times. Hence, the disk-based algorithms are designed mainly to maximize the main memory utilization and the disk access efficiency. However, these algorithms have a common drawback that they incur random disk accesses. The disk access performance is dependent more on access patterns than access amount; even for accessing the same amount, the random disk access requires much more time than the sequential disk access. Thus, the disk-based algorithms have been improved in the way of decreasing the ratio of random disk accesses.

Another problem of the previous disk-based algorithms is that they do not fully utilize the most up-to-date CPU technologies. Instead of raising the clock speed, recent CPUs are designed to have multiple, simultaneously running cores that enable intra-CPU parallel processing. However, some previous algorithms run mostly on a single core, and the others suffer from severe interference among the threads and hence have little gain by parallel processing. We explain the problems of the previous algorithms in more detail in Section~\ref{sec:related}.

In this paper, we propose a fast algorithm based on `divide-and-conquer' strategy for constructing the suffix tree for large-scale human genome sequences. The most significant difference from the previous algorithms is that the proposed algorithm almost eliminates random disk accesses by accessing the disk in the unit of contiguous chunks each of which stores an entire suffix subtree. In addition, our algorithm fully utilizes the multi-core CPUs by dividing the genome sequences into multiple, independent partitions and then assigning each partition to a different core for parallel construction of suffix subtrees. 
As an experimental result, our algorithm finished construction of the suffix tree for the entire human genome sequences in 64 minutes and outperformed DIGEST algorithm~\cite{bar08}, which had previously been the fastest disk-based algorithm, by up to 3.5 times.

This paper is organized as the following. In Section~\ref{sec:suffix}, we briefly explain on the suffix tree. In Section~\ref{sec:related}, we explain on the previous disk-based suffix tree construction algorithms. We also explain the performance degradation by random disk accesses in the section. In Section~\ref{sec:indexing}, we propose a new disk-based suffix tree construction algorithm, and then in Section~\ref{sec:eval}, we evaluate the performance of our algorithm through a series of experiments.

\section{Suffix tree}
\label{sec:suffix}

\begin{figure}[t]
\centering
\includegraphics[width=2.5in]{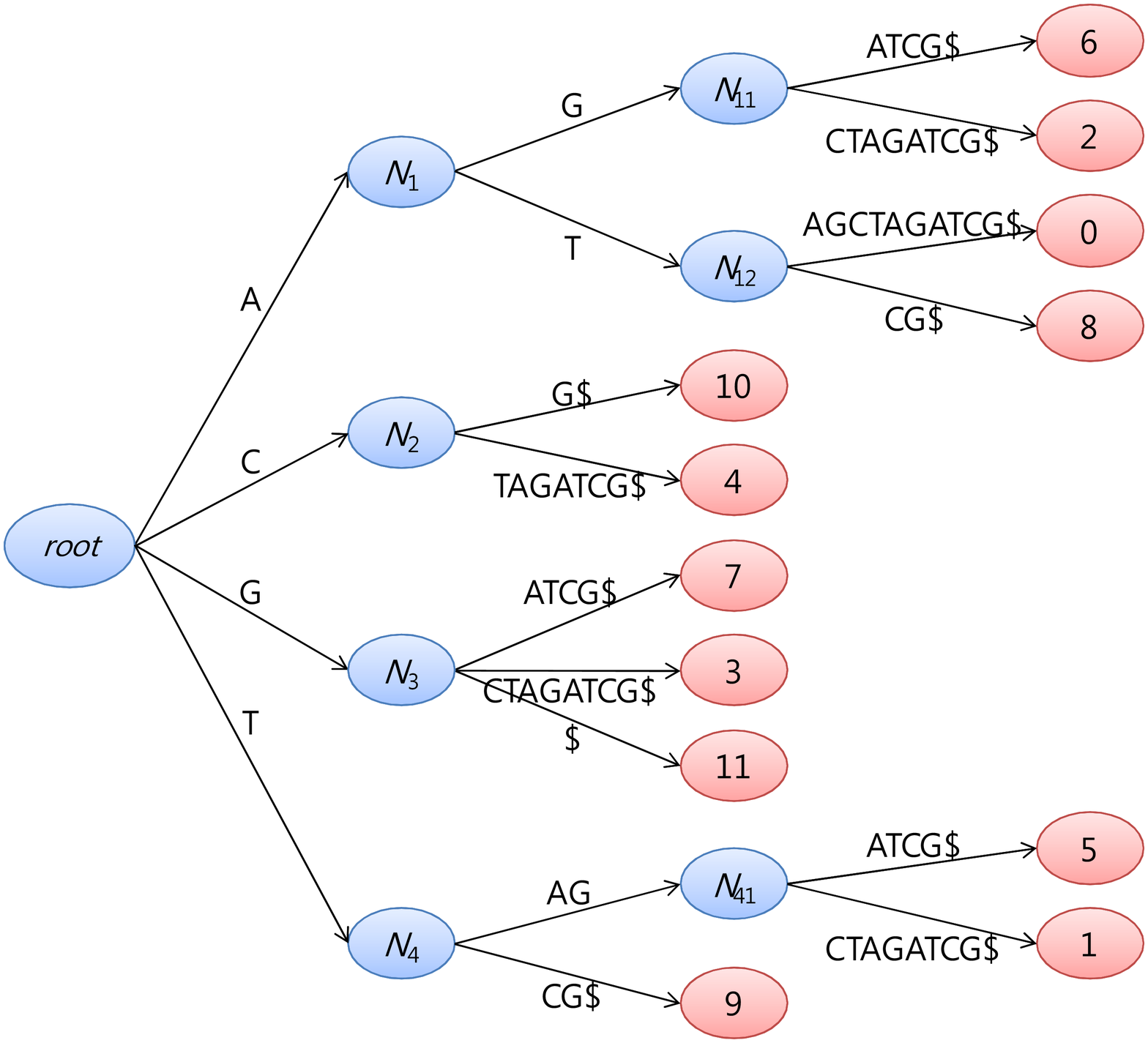}\\
{\footnotesize (a) Edge labels are represented with subsequences.}\\[0.1in]
\includegraphics[width=2.5in]{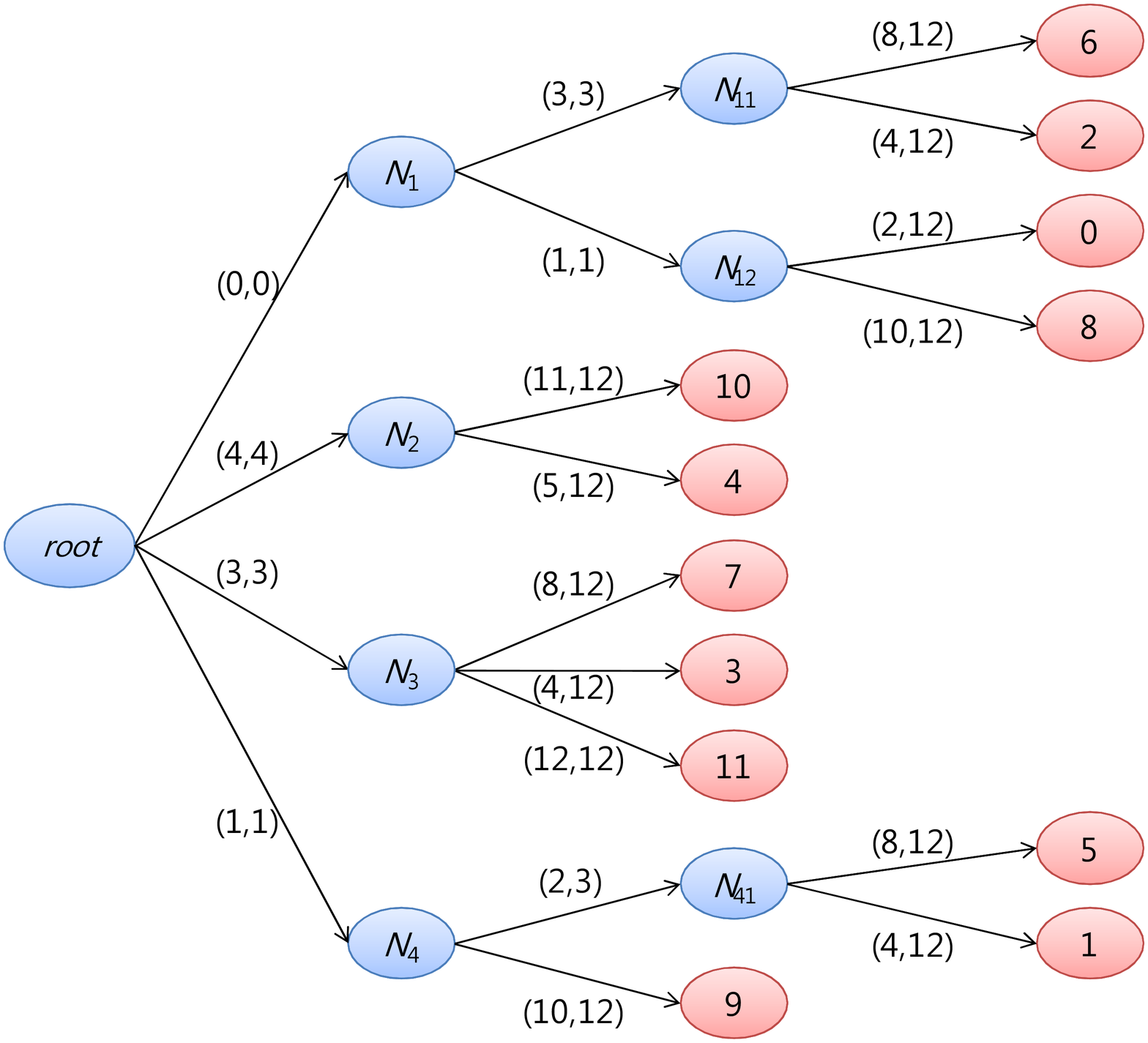}\\
{\footnotesize (b) Edge labels are represented with (start, end) positions in $X$.}
\caption{Suffix tree for a sequence $X$ = ATAGCTAGATCG\$.}
\label{fig01}
\end{figure}

Figure~\ref{fig01} shows the suffix tree for a short DNA sequence $X$ = ATAGCTAGATCG\$. The symbol `\$' is appended at the end of $X$ so as to prohibit any suffix in $X$ from being the prefix of any other suffix. Given a query sequence $S$, the search begins from the root node of the suffix tree. From the outbound edges of the root node, an edge $e$ is chosen such that the label of $e$ is the prefix of $S$. If no such edge is found, the search ends; if found, the child node $N_e$ is visited by following the edge $e$, i.e., $e$ is the inbound edge of $N_e$. Let $l$ be the label length of $e$, $p_l(S)$ be the prefix of $S$ of length $l$, and $s_l(S)$ be the suffix of $S$ of length $Len(S) - l$. Then, it holds that $S = p_l(S) \oplus s_l(S)$, where $\oplus$ is the sequence concatenation operator. The search for query subsequence $s_l(S)$ begins recursively at the node $N_e$ in the same manner as the root node. The search goes on until a terminal node is reached in the suffix tree or there is no query (sub)sequence to be searched for.

Let us take a query sequence $S$ = AGATCG for example. In Figure~\ref{fig01}(a), from the outbound edges of the root node, the edge with label `A' is followed and then the node $N_1$ is visited. The search for query subsequence $s_l(S)$ = GATCG is performed recursively at the node $N_1$. The search continues until the terminal node with position 6 is reached; it indicates that query sequence $S$ is found at position 6 in the sequence $X$. Figure~\ref{fig01}(b) shows the suffix tree whose edge labels are represented with (start, end) positions in $X$. While the labels' representation sizes in Figure~\ref{fig01}(a) are arbitrary, those in Figure~\ref{fig01}(b) are all identical.

\section{Related work}
\label{sec:related}

Hunt et al.~\cite{hun02} proposed the first disk-based suffix tree construction algorithm. Hunt's algorithm excludes construction of suffix links, which caused severe memory bottleneck problem in Ukkonen's algorithm~\cite{ukk95}. Hunt's algorithm divides the given genome sequences into partitions and then constructs a separate suffix subtree for each partition. Although Hunt's algorithm has $O(n^2)$ complexity, it shows better indexing performance than Ukkonen's algorithm by reducing disk accesses. However, Hunt's algorithm incurs heavy random disk accesses since it stores each node in the suffix tree as a separate object using the persistent Java object storage interface called PJama~\cite{atk98}. Actually, the algorithm was successful in indexing genome sequences of up to 286Mbp size, but it could not be used for indexing the human genome sequences~\cite{hun02}.

Tian et al.~\cite{tia05} presented the Top-Down Disk-based (TDD) approach for constructing disk-based suffix trees. TDD consists of two algorithms: Partition and Write Only Top Down (PWOTD) algorithm based on Wotd-eager algorithm~\cite{gie03} for constructing suffix trees and a memory buffer management algorithm for maximizing the performance of PWOTD algorithm. The performance of PWOTD algorithm highly depend on the settings of the memory buffer management algorithm~\cite{tia05}. Tian et al.~\cite{tia05} showed that TDD incurred only one sixth of disk accesses than DynaCluster algorithm~\cite{che05}, an extension of Hunt's algorithm, and that TDD constructed the suffix tree for the entire human genome sequences in 30 hours. However, the memory buffer management algorithm in TDD assigns only a small portion of memory for keeping the suffix tree in main memory, while it assigns the largest portion to input genome sequences. TDD uses Least Recently Used (LRU) policy for swapping out the memory buffers into disk while constructing the suffix tree. Whenever PWOTD algorithm creates a new node $N$, it needs to access $N$'s parent node $P$ that could be previously stored far away from $N$. This causes random disk accesses, and the larger genome sequences should cause more random accesses.

Phoophakdee and Zaki~\cite{pho07} proposed an algorithm called TRELLIS, which eliminated data skewness among suffix subtrees by dividing genome sequences according to variable-length prefixes. Unlike Hunt's algorithm~\cite{hun02} and TDD~\cite{tia05}, TRELLIS can create suffix links optionally after the suffix tree is constructed. TRELLIS consists of three phases: prefix creation, partitioning, and merging phases. In the prefix creation phase, variable-length prefixes are created so that, for each prefix $P_j$, the suffix subtree $T_j$ corresponding to the suffixes having the prefix $P_j$ can be loaded into main memory as a whole. In the partitioning phase, the entire genome sequences are divided into partitions so that each partition $R_i$ and its corresponding suffix tree $T_i$ can be loaded into main memory as a whole. Then, a suffix tree $T_i$ is constructed for each partition in this phase. In the merging phase, for each prefix $P_j$ created in the prefix creation phase, the suffix subtrees $T_{i,j}$ are extracted from the suffix trees $T_i$ and then merged into a single suffix subtree $T_j$. Phoophakdee and Zaki~\cite{pho07} showed that TRELLIS outperformed TDD by up to 4 times and that it constructed the suffix tree for the entire human genome sequences in 4.2 hours. However, since TRELLIS extracts the suffix subtrees $T_{i,j}$ stored at random positions in the suffix trees $T_i$ in the merging phase, it incurs severe random disk accesses. Actually, the merging phase requires the longest execution time~\cite{pho07}.

Ghoting and Makarychev~\cite{gho09} proposed an algorithm called WAVEFRONT based on `partition-and-merge' strategy as TRELLIS~\cite{pho07}. WAVEFRONT divides the entire data into I/O-efficient partitions and processes each partition independently. In~\cite{gho09}, WAVEFRONT was extended to be executed on a massively parallel system. The algorithm completed indexing the entire human genome sequences in 15 minutes on IBM Blue Gene/L system composed of 1024 processors~\cite{gho09}. However, WAVEFRONT executed on a single processor showed no noticeable performance improvement compared with TRELLIS~\cite{pho07}.

Barsky et al.~\cite{bar08} proposed an algorithm called DIGEST which consists of two phases similar to the merge-sort algorithm. In the first phase, the entire genome sequence is divided into partitions of the same length so that each partition can be loaded into main memory. For each partition, the suffixes contained therein are sorted in main memory and then are stored in disk. In the second phase, the suffixes sorted separately in each partition are merge-sorted. Suffix blocks from each partition are read sequentially one by one into main memory. The suffixes in different blocks are compared with each other, and the smallest one is extracted and then saved in the output block. When the output block becomes full, it is stored in disk. This continues until all the input blocks are empty. The sorted suffixes is called a {\it suffix array\/}, and it is known that a suffix array can be easily converted into a suffix tree~\cite{bar08, sin08}. Barsky et al.~\cite{bar08} showed that DIGEST outperformed TRELLIS+~\cite{pho08}, an extension of TRELLIS~\cite{pho07}, by up to 40\% and that the algorithm completed indexing the entire human genome sequences in about 85 minutes. However, DIGEST should read suffix blocks from each partition stored at random positions in the second phase and hence suffers from severe random disk accesses. Moreover, since the merging phases of TRELLIS and DIGEST cannot be parallelized, they have little performance gain even by using recent multi-core CPUs.

\begin{figure}
\centering
\includegraphics[width=3.2in]{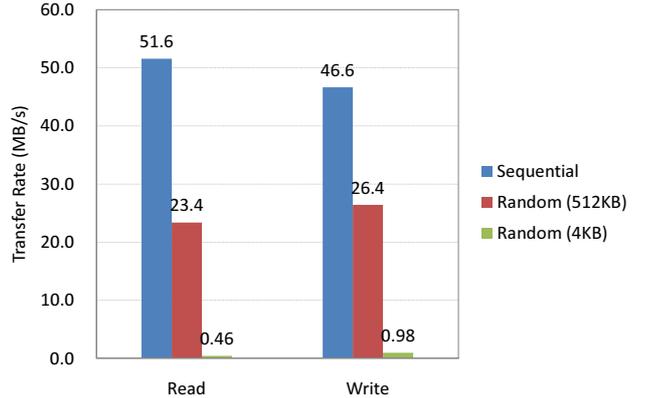}
\caption{Disk read/write transfer rates: sequential read/write performed much better than random read/write.}
\label{fig02}
\end{figure}

As explained so far, the common drawback of the previous algorithms is the performance degradation due to random disk accesses. Figure~\ref{fig02} shows an experimental result of reading/writing a disk volume of 100MB size. The volume was read and written sequentially and at random in the unit of 512KB and 4KB. In the figure, the sequential read/write performed up to 112.1 and 47.7 times better than random read/write, respectively. The values in Figure~\ref{fig02} should be different according to experimental environments, though it is always the case that sequential accesses have better performance than random accesses.

\section{Proposed indexing algorithm}
\label{sec:indexing}

In this section, we propose a new algorithm for indexing human genome sequences. The human genome is composed of 46 chromosomes: 22 chromosome pairs numbered 1 $\sim$ 22 and x/y (sex) chromosomes. In this paper, we concatenate the entire genome sequences into a single long sequence and use this sequence as the input of our algorithm. This helps simplify indexing and searching algorithms. 

Our algorithm is designed based on divide-and-conquer strategy: it divides the entire human genome sequence into multiple independent partitions and then constructs the suffix tree separately for each partition. The suffix tree for each partition is constructed in a contiguous chunk in main memory. When the construction is completed, the chunk image is stored sequentially into disk as it is. Hence, unlike TRELLIS and DIGEST~\cite{bar08, pho07}, our algorithm has no performance degradation due to random disk accesses. Moreover, since the suffix trees for different partitions are constructed independently and are not merged thereafter, their construction can be done in parallel by fully utilizing the most up-to-date multi-core CPUs. According to these features, our algorithm achieves dramatic performance improvement compared with the previous algorithms.

Our algorithm represents each base as a 2-bit code as in~\cite{bar08, pho07, pho08, won07}; A, C, G, and T are represented as 00, 01, 10, and 11, respectively. Since the human genome sequence has the size of approximately 3Gbp, the 2-bit coded sequence has the size of about 3Gbp / 4 = 750MB. Actually, after removing unidentified base pairs, the 2-bit coded sequence has the size of about 700MB and can be fully loaded in main memory. Our algorithm assigns memory region for the full 2-bit coded genome sequence at the beginning and retains it to the end.

Our algorithm divides the human genome sequence into partitions according to prefixes, i.e., the suffixes having the common prefix belong to the same partition. We explain how to determine the prefixes for partitioning at the end of this section. The partitions are not necessarily created by physically dividing the genome sequence, but only the suffix positions are managed for each partition. While scanning the entire genome sequence, our algorithm creates the lists of suffix positions simultaneously for every prefix determined earlier; the list for a prefix $P_j ~ (0 \leq j < m)$ contains the positions of suffixes having the prefix $P_j$, where $m$ is the number of partitions. Although each of these lists has a small size, the entire lists occupy a considerable amount of memory. Hence, the lists are stored in disk right after their creation; each list is retrieved from disk only once when the suffix tree is about to be constructed for the corresponding partition.
Our algorithm creates each list of suffix positions in a contiguous memory region to read/write the list with a single operation and hence to eliminate random disk accesses. To obtain the sizes of contiguous memory regions, our algorithm scans the human genome sequence to count the frequency of every prefix before creating the lists of suffix positions.

\begin{figure}[t]
\centering
\begin{tabular}{ccc}
\includegraphics[scale=0.3]{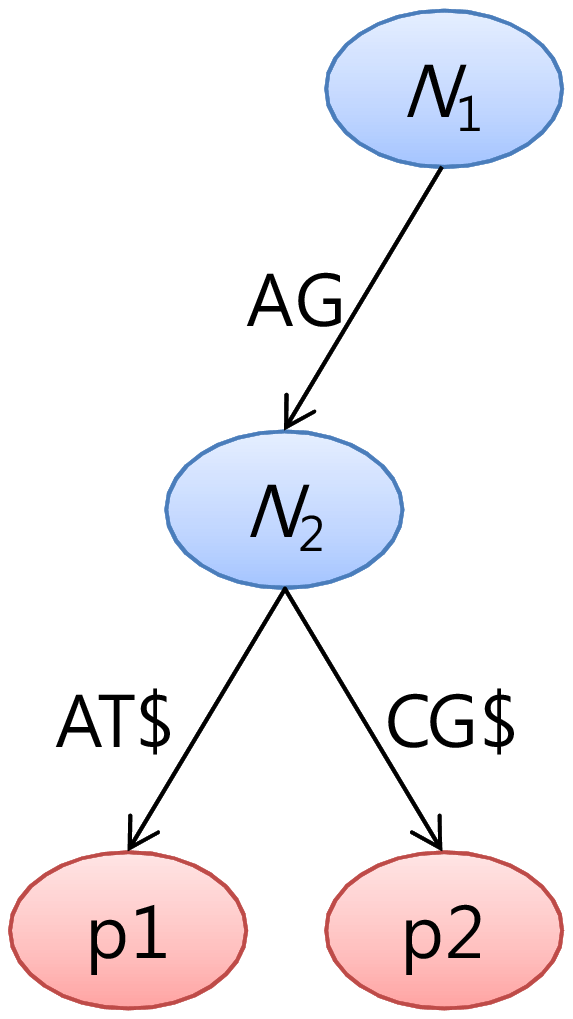} &
\includegraphics[scale=0.3]{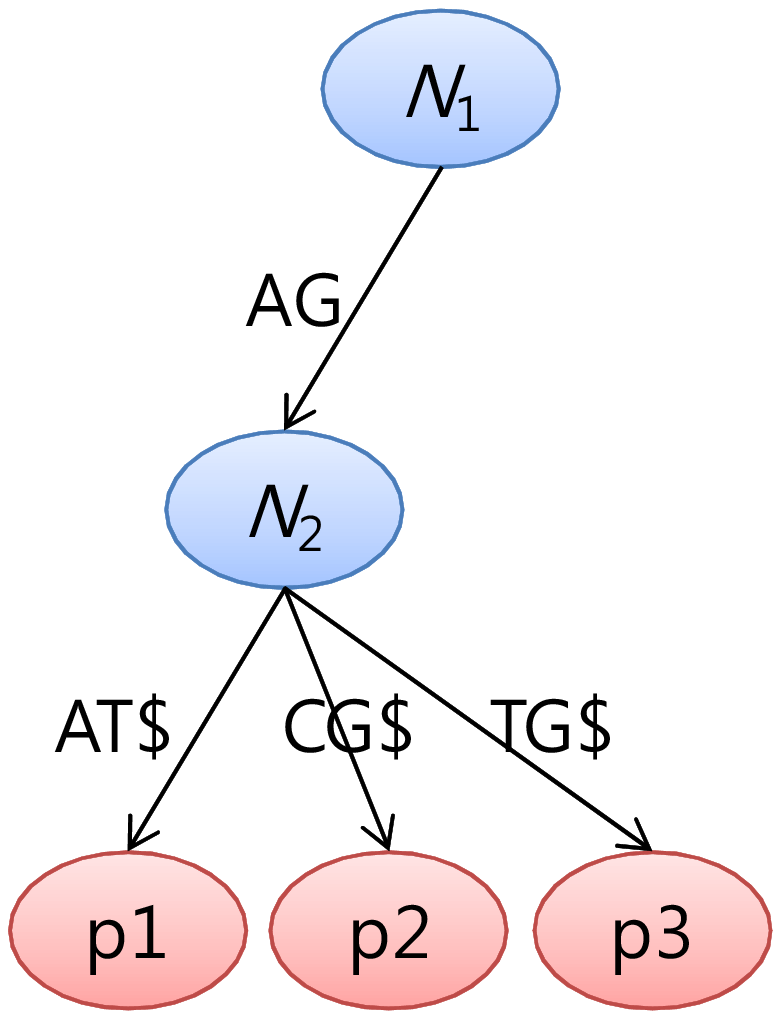} &
\includegraphics[scale=0.3]{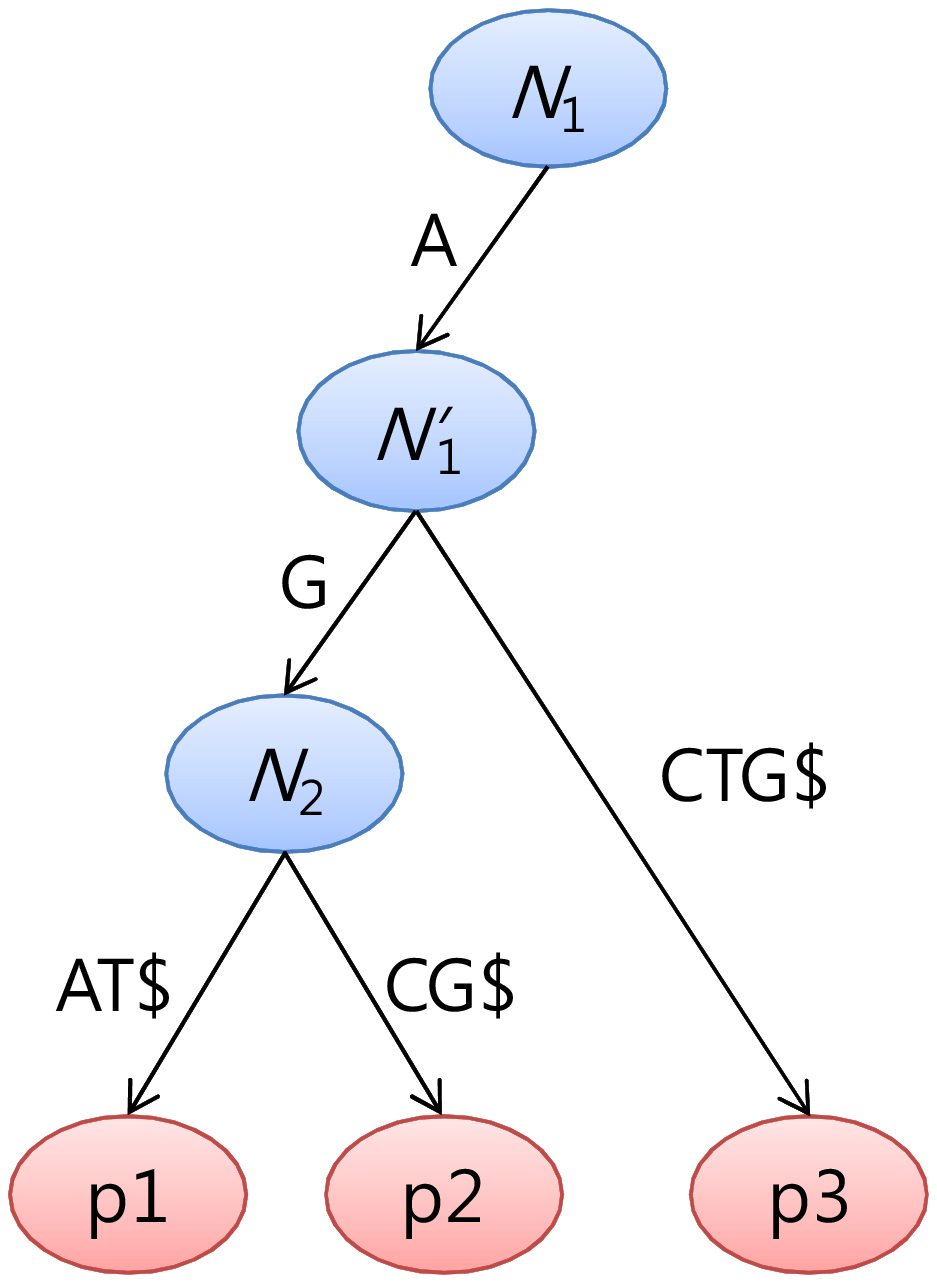}\\
\parbox{0.8in}{\footnotesize (a) A suffix tree.} &
\parbox{1.0in}{\footnotesize (b) Adding a suffix $S_1$ = AGTG\$.} &
\parbox{1.2in}{\footnotesize (c) Adding a suffix $S_2$ = ACTG\$.}\\
\end{tabular}
\caption{Example of adding suffixes into a suffix tree.}
\label{fig03}
\end{figure}

When the creation of partitions (i.e., the lists of suffix positions in the human genome sequence) is completed, our algorithm constructs the suffix tree separately for each partition. At first, our algorithm creates an empty suffix tree without any node and then adds suffixes one by one into the suffix tree while scanning the corresponding list of suffix positions. Figure~\ref{fig03} shows an example of adding suffixes into a suffix tree. Figure~\ref{fig03}(a) shows a suffix tree before addition. Figure~\ref{fig03}(b) shows the result of adding a suffix $S_1$ = AGTG\$ into the suffix tree in Figure~\ref{fig03}(a). $S_1$ has the prefix $p_2(S_1)$ = AG of length 2 which matches the label of the outbound edge of $N_1$ and then $s_2(S_1)$ = TG\$ does not have common prefix with any label of the outbound edges of $N_2$. In this case, our algorithm creates a new outbound edge $e$ of $N_2$ and labels it with $s_2(S_1)$ = TG\$. The edge $e$ is connected to a new terminal node $p_3$, i.e., $e$ becomes the inbound edge of $p_3$. Figure~\ref{fig03}(c) shows the result of adding a suffix $S_2$ = ACTG\$ into the suffix tree in Figure~\ref{fig03}(a). The label of the outbound edge of $N_1$ partially matches the prefix $p_1(S_2)$ = A of $S_2$. In this case, our algorithm cuts the outbound edge of $N_1$ and adds a new internal node $N'_1$; the inbound edge of $N'_1$ has the label $p_1(S_2)$ = A. A new outbound edge $e$ is added to node $N'_1$ and is labeled with $s_1(S_2)$ = CTG\$. The edge $e$ is connected to a new terminal node $p_3$, i.e., $e$ becomes the inbound edge of $p_3$.

Each time a suffix is added into the suffix tree, a new terminal node is created in the tree. Since every suffix ends with the symbol \$, the suffix cannot be a prefix of any other suffixes and has a unique position in the human genome sequence. Hence, a terminal node should exist in the suffix tree for representing the unique position of each suffix. The terminal node should have an inbound edge in the tree. The edge is an outbound edge of either (1)~an existing node (Figure~\ref{fig03}(b) case) or (2)~a new node added between the cut edges (Figure~\ref{fig03}(c) case). There exist no other cases.

\begin{figure}
\centering
\begin{tabular}{ccc}
\includegraphics[scale=0.3]{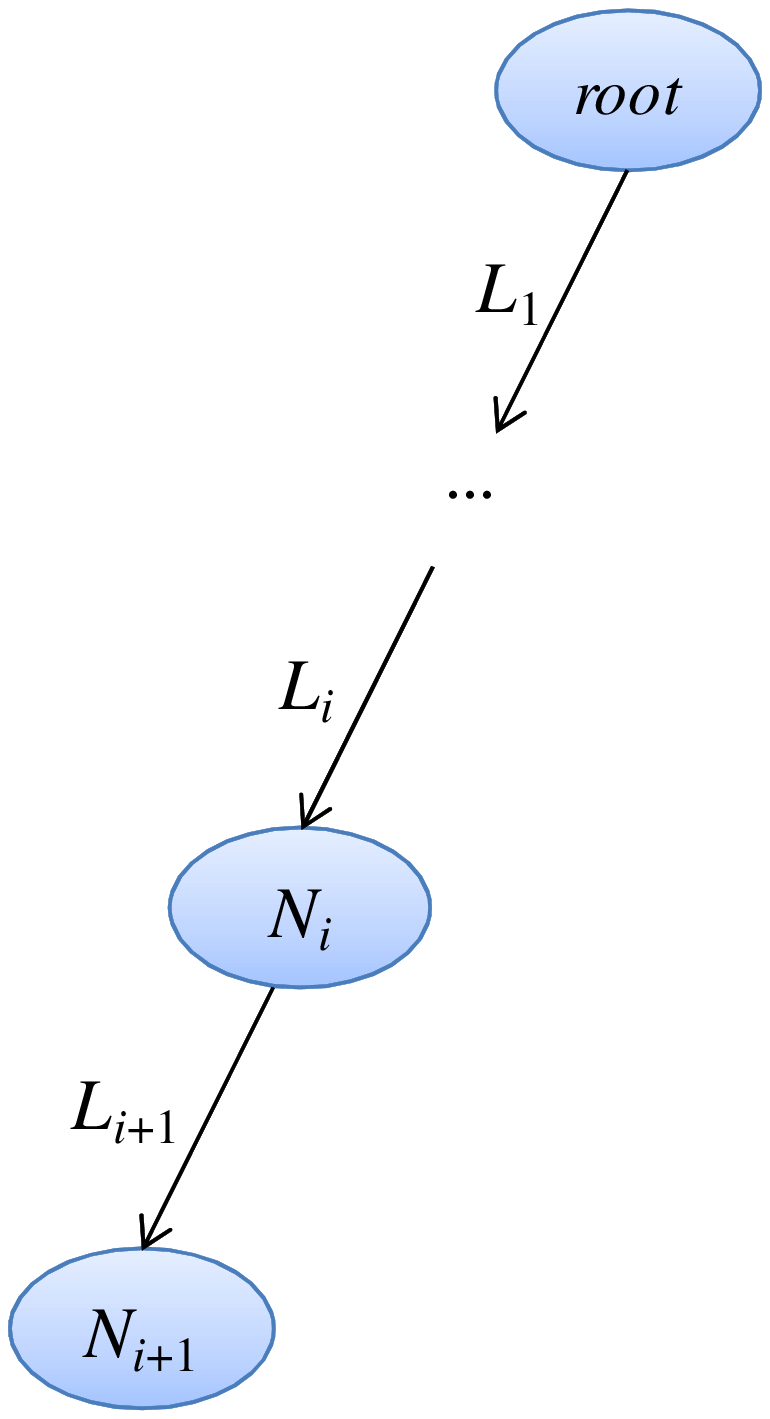} &
\includegraphics[scale=0.3]{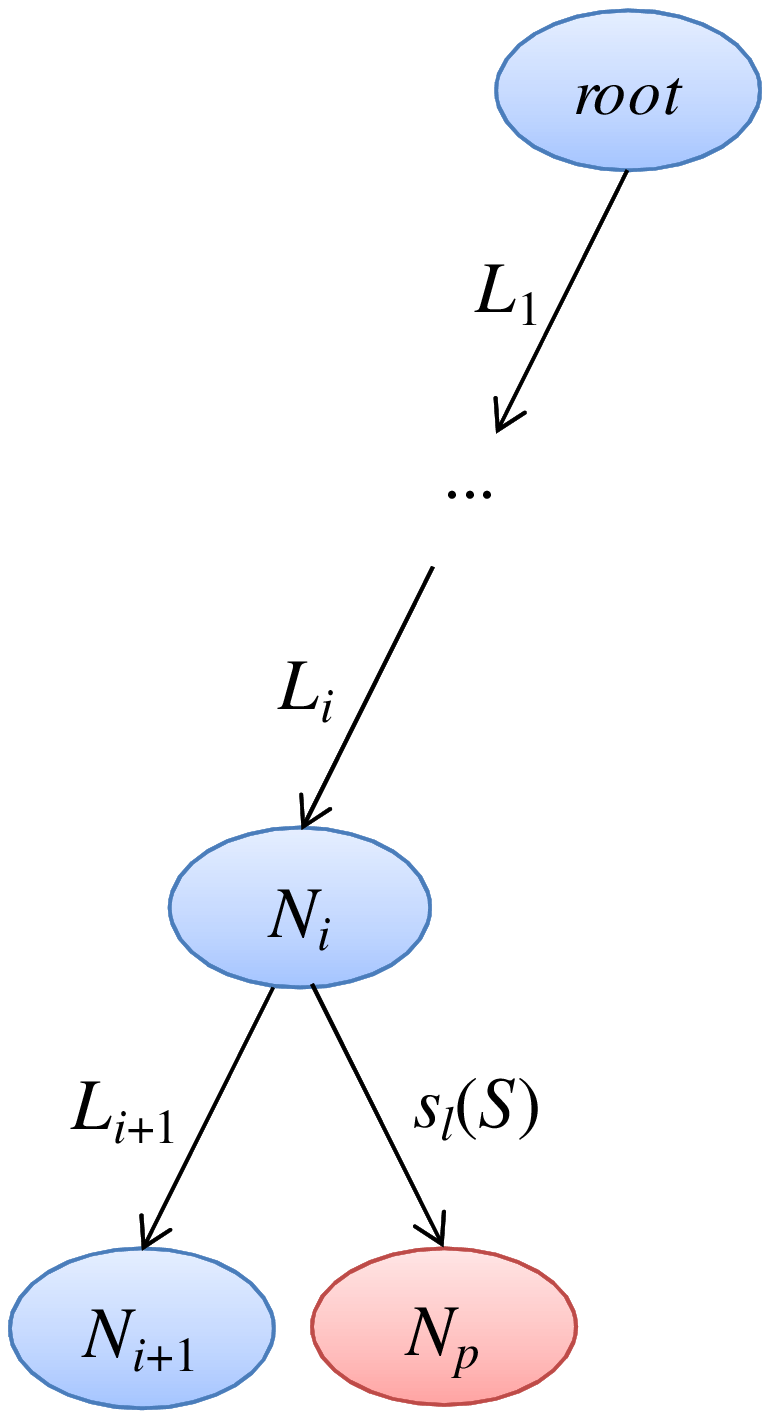} &
\includegraphics[scale=0.3]{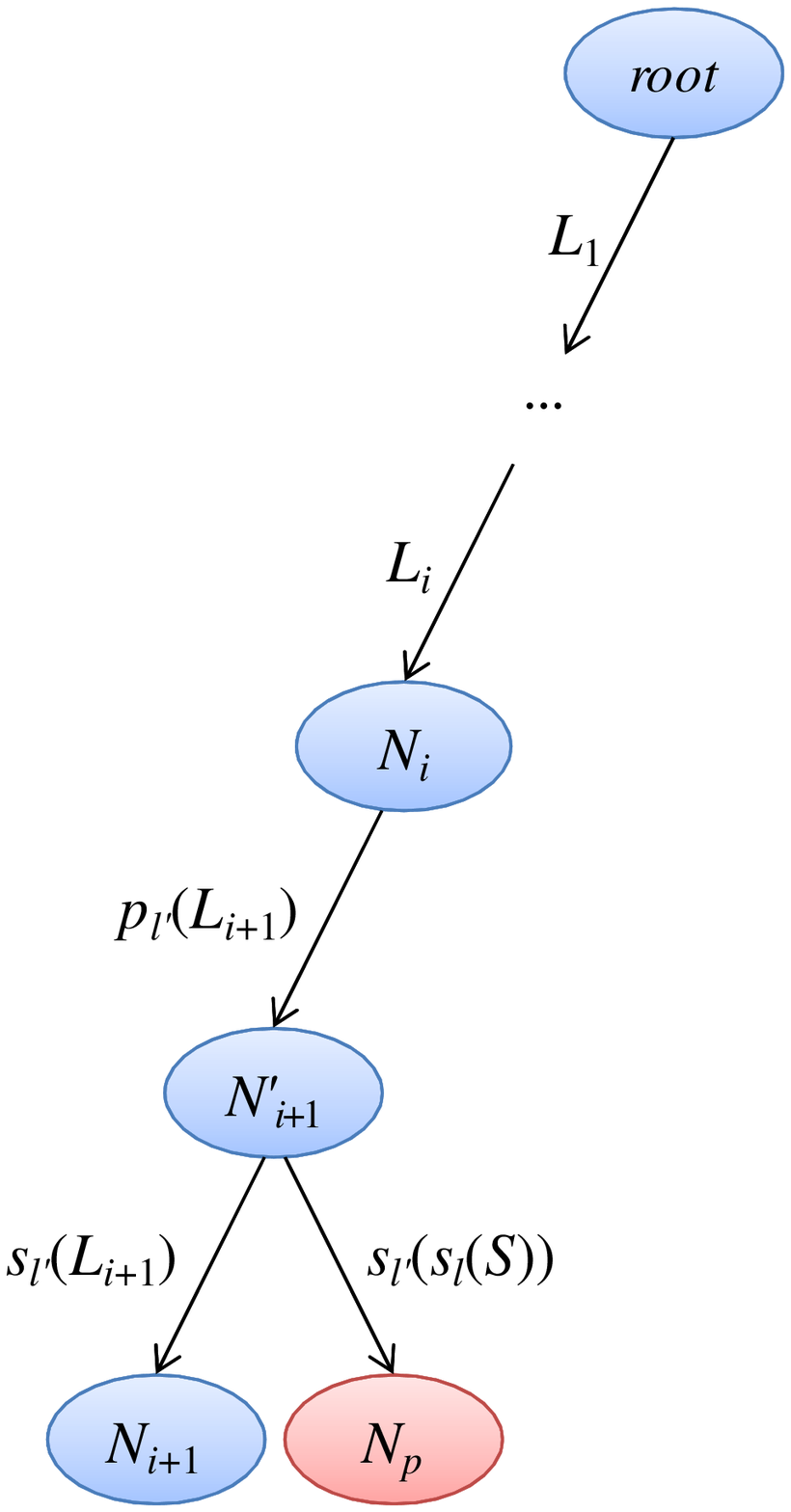}\\
\parbox{0.95in}{\footnotesize (a) Before adding a suffix.} &
\parbox{0.95in}{\footnotesize (b) The case of adding a terminal node.} &
\parbox{1.10in}{\footnotesize (c) The case of adding an internal node and a terminal node.}\\
\end{tabular}
\caption{Generalization of adding suffixes into a suffix tree.}
\label{fig04}
\end{figure}

Figure~\ref{fig04} shows the generalization of adding suffixes into the suffix tree by our algorithm. Let us assume that we have visited the node $N_i$ in the course of searching for a suffix $S$ in Figure~\ref{fig04}(a). The concatenation $L = L_1 \oplus \dots \oplus L_i$ of edge labels from the root node to $N_i$ should be the same as the prefix $p_l(S)$ of length $l = Len(L)$, i.e., $L = p_l(S)$. In case $L_{i+1} \cap s_l(S) = \varnothing$, an edge $e$ labeled with $s_l(S)$ and a new terminal node $N_p$ with the inbound edge $e$ are added as in Figure~\ref{fig04}(b). In case $L_{i+1} \cap s_l(S) = L' ~ (\neq \varnothing)$, a new internal node $N'_{i+1}$ and a new terminal node $N_p$ are added as in Figure~\ref{fig04}(c), where $l' = Len(L')$ and $p_{l'} \left ( L_{i+1} \right ) = p_{l'} \left ( s_l(S) \right ) = L'$. Since the suffix always ends with \$, we cannot have the case $S = L$ in Figure~\ref{fig04}.

\begin{figure}
\centering
\includegraphics[width=3.5in]{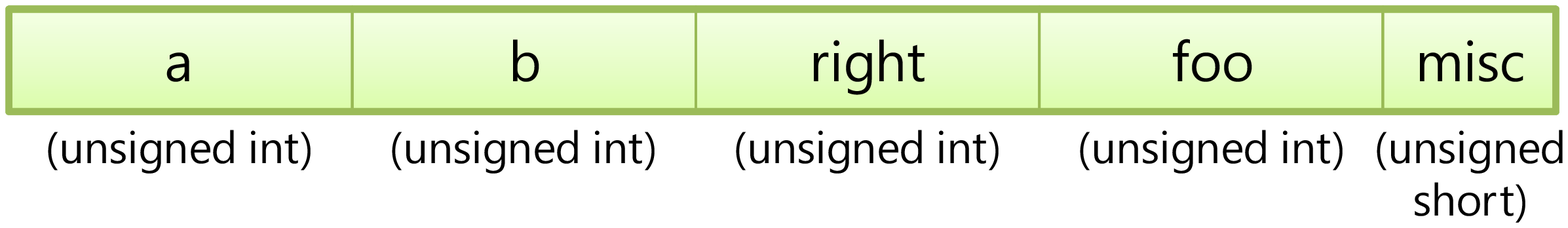}
\caption{Data structure of our algorithm: the information on a node and its inbound edge is contained together.}
\label{fig05}
\end{figure}

Figure~\ref{fig05} shows the data structure of our algorithm. As shown in the figure, the information on a node and its inbound edge is contained together in a single data structure. The fields $a$ and $b$ represent the start and end positions of the inbound edge in the human genome sequence as shown in Figure~\ref{fig01}(b). The field {\it right\/} contains the pointer to the next sibling node, and {\it foo\/} represents either (1)~a pointer to the leftmost child node in case of an internal node or (2)~the suffix position in the genome sequence in case of a terminal node. The field {\it misc\/} contains miscellaneous information on the node. The fields $a$, $b$, {\it right\/}, and {\it foo\/} are 4-byte unsigned integers, while the field {\it misc\/} is a 2-byte unsigned integer. Hence, the data structure has the fixed length of 18 bytes. For distinguishing between the internal and terminal nodes, the field $b$ is investigated. If $b = n$, where $n$ is the length of genome sequence, it is a terminal node; if $b < n$, it is an internal node (refer to Figure~\ref{fig01}(b)). 

\begin{figure*}
\centering
\includegraphics[scale=0.4]{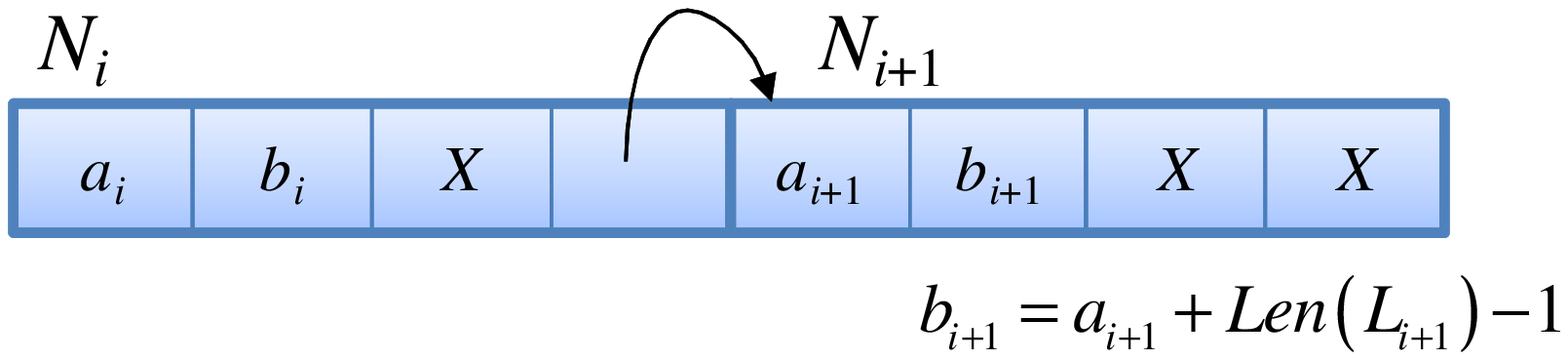}\\
{\footnotesize (a) Before adding a suffix.}\\[0.1in]
\includegraphics[scale=0.4]{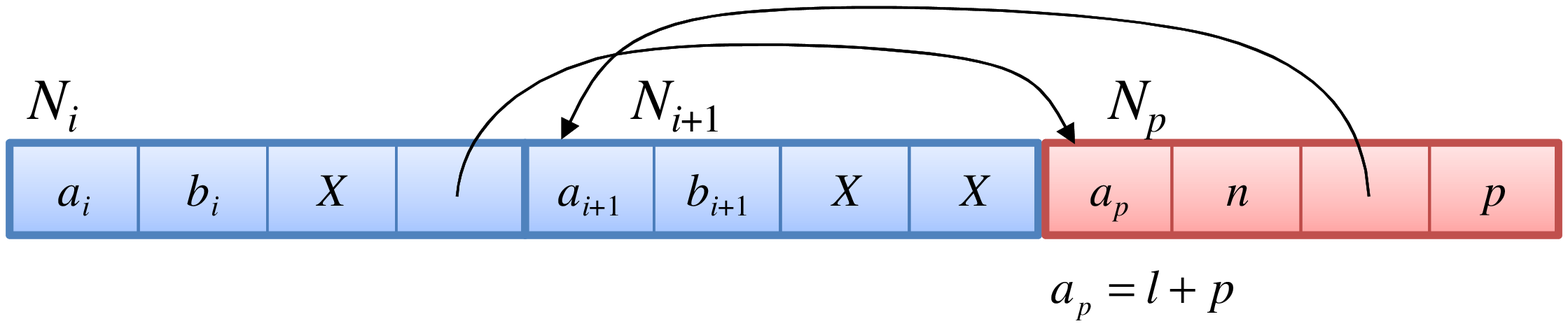}\\
{\footnotesize (b) The case of adding a terminal node.}\\[0.1in]
\includegraphics[scale=0.4]{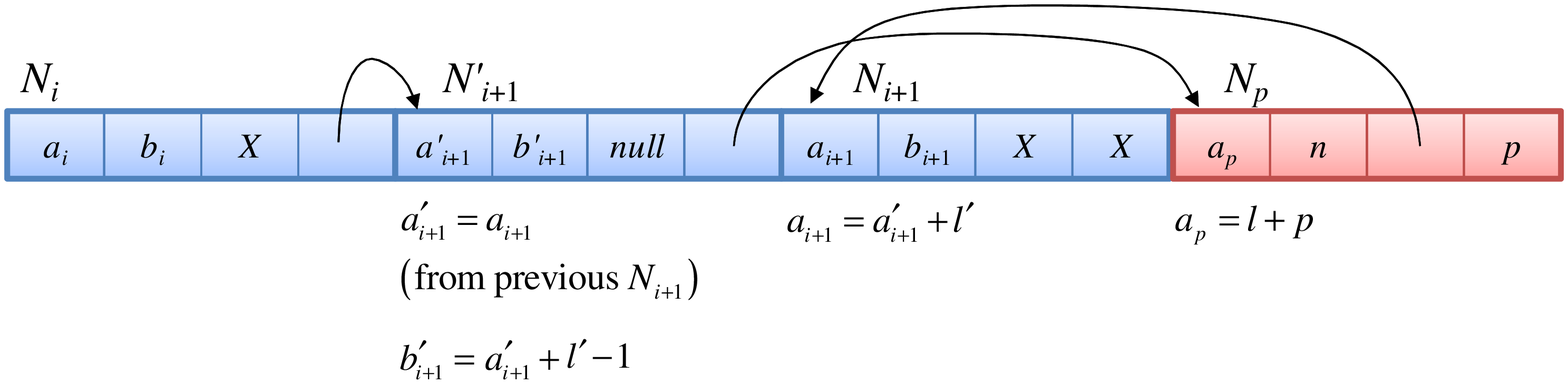}\\
{\footnotesize (c) The case of adding an internal node and a terminal node.}\\
\caption{Data structures corresponding to the suffix trees in Figure~\ref{fig04}.}
\label{fig06}
\end{figure*}

We can efficiently construct the suffix trees using the data structure in Figure~\ref{fig05}. We explain this using Figure~\ref{fig06}, which shows the representation of suffix trees in Figure~\ref{fig04} using the data structure; Figures~\ref{fig06}(a) $\sim$ \ref{fig06}(c) correspond to Figures~\ref{fig04}(a) $\sim$ \ref{fig04}(c), respectively. In Figure~\ref{fig06}(a), the fields $(a_i, b_i)$ and $(a_{i+1}, b_{i+1})$ represent the start and end positions of labels $L_i$ and $L_{i+ 1}$, respectively. The fields with $X$ stand for ``don't care'' fields, which are not used nor updated here. The arrow indicates a pointer to a possible distant node. The nodes $N_i$ and $N_{i+1}$ may not be adjacent as shown in the figure, though $N_{i+1}$ is easily accessed by following the pointer. Figure~\ref{fig06}(b) shows the case a new terminal node $N_p$ is added. The node $N_{i+1}$ can be either an internal or a terminal node and is a sibling node of $N_p$. In the figure, the leftmost child node of $N_i$ has been changed from $N_{i+1}$ to $N_p$. This is because we can efficiently add $N_p$ as a new child node of $N_i$ without accessing $N_{i+1}$ and all its sibling nodes. Figure~\ref{fig06}(c) shows the case a new internal node $N'_{i+1}$ and a new terminal node $N_p$ are added. The field values of the $N_{i+1}$ are copied to the newly allocated node region, and then the field $a_{i+1}$ is adjusted ($b_{i+1}$ is not changed). The field values of $N'_{i+1}$ are set in the region previously used by $N_{i+1}$ as shown in the figure. The node $N_p$ is a sibling node of $N_{i+1}$ and is added as the leftmost child node of $N'_{i+1}$ as in Figure~\ref{fig06}(b). The key idea we would like to show in Figure~\ref{fig06} is that, when a suffix is added, there is only slight modification in the suffix tree constructed so far; it can be done only by allocating new memory region(s) for one or two nodes and then setting a few appropriate field values therein. This is one of the features providing the efficiency of our algorithm.

Our algorithm constructs a suffix tree in a main memory chunk. Allocations of memory regions for new nodes (and their inbound edges) are made sequentially in the chunk. The pointers in Figures~\ref{fig05} and \ref{fig06} are relative offset values from the beginning of the chunk. Once the construction of a suffix tree is completed, our algorithm stores the chunk image into disk without any modification. When the chunk image is reloaded into main memory, the pointers are still valid regardless of where it is reloaded. Since the chunk image is stored in and read from the disk sequentially, there is no performance degradation due to random disk accesses, and thus we have significantly improved performance. When multiple suffix trees are constructed in parallel, our algorithm allocates a separate memory chunk for each suffix tree. Even in this case, the human genome sequence is loaded only once into the memory region shared by the simultaneous processes of our algorithm. This parallel processing enables more significant performance improvement.

We now explain how to determine the prefixes for dividing the human genome sequence into partitions. Each suffix in the genome sequence is assigned to a partition according to its prefix; every suffix in a partition has a common prefix. Given a prefix length $p$, our algorithm creates a partition for each possible prefix of length $p$. The number of partitions is $4^p$. A weakness of this scheme is that it causes data skewness among the partitions~\cite{pho07}; there may be big differences among the sizes of partitions and hence the corresponding suffix trees. We tackle this weakness as follows. As $p$ increases, the number of suffixes in each partition decreases, and the size of corresponding suffix tree also decreases. We set $p$ to be large enough to make the suffix tree sizes smaller than the size $M$ of available main memory. Then, the simultaneous processes of our algorithm choose the partitions so that the estimated sizes of their corresponding suffix trees sum up very close to $M$. This can be done with simple computations. By fully utilizing main memory in this way, our algorithm achieves better indexing performance.

The minimum length of prefixes is computed approximately using the following Eq.~(\ref{eq1}):
\begin{eqnarray}
p_{\min} = \left \lceil \log _4 \frac{n \cdot f}{M} \right \rceil ~ , \label{eq1}
\end{eqnarray}
where $n$ is the length of human genome sequence and $f$ is a multiplication factor to estimate the suffix tree size. $M$ represents the size of remaining main memory after loading the entire 2-bit coded human genome sequence. $f$ is defined as the maximum of $\frac{T}{s}$, where $s$ is the length of a genome sequence and $T$ is the size of the corresponding suffix tree. We estimate the size of a big suffix tree by test construction of small suffix trees. The $f$ value greatly differs according to suffix tree construction algorithms and is about 30 $\sim$ 32 in our algorithm.

\section{Performance evaluation}
\label{sec:eval}

In this section, we show the superiority of our algorithm through a series of experiments. We use the same data sets as those in~\cite{bar08}. The first set is a short genome sequence of 110Mbp size obtained from 6643 organisms. The second set is the entire human genome sequence of about 3Gbp size. These data sets are denoted as VDB and HG18, respectively.

The hardware platform is a PC equipped with Intel Core2Quad Q9550 2.83GHz CPU, Samsung DDR3 8GB main memory, and a 500GB 7200rpm hard disk. The software platforms are Ubuntu 10.10 32bit Linux and Windows 7 64bit Edition. The first experiment was performed on Ubuntu as in~\cite{bar08}, and the second and third experiments were performed on Windows 7. The latter two experiments were also performed on Ubuntu, though we had 10 $\sim$ 15\% better performance on Windows 7. As C/C++ compilers, we used GNU C++ 4.4.5 on Ubuntu and Visual C++ 2010 Express Edition on Windows 7.

\begin{figure}
\centering
\includegraphics[width=3.0in]{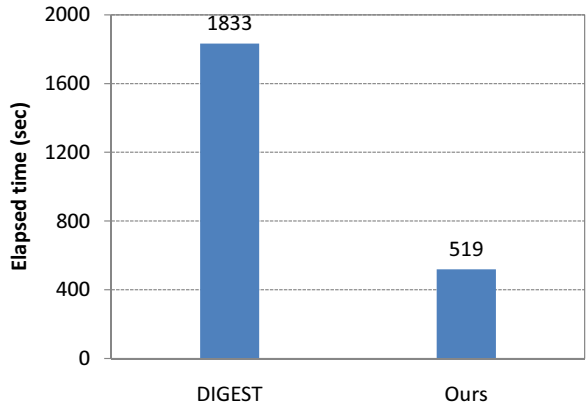}
\caption{Result of first experiment: our algorithm outperformed DIGEST by up to 3.5 times.}
\label{exp1}
\end{figure}

In the first experiment, we compared the performance of our algorithm with DIGEST~\cite{bar08}, which had been the fastest disk-based suffix tree construction algorithm. We downloaded the source code of DIGEST from the author's web site\footnote{\url{http://webhome.cs.uvic.ca/~mgbarsky/}}. In this experiment, we ran our algorithm and DIGEST on VDB data set and compared their elapsed time for constructing the suffix trees\footnote{We also tried the experiment on HG18 data set; however, DIGEST always terminated abnormally with the segmentation fault error. We discussed on this with the author of DIGEST, but we could not solve the problem to the end.}. Figure~\ref{exp1} shows the result of experiment; our algorithm outperformed DIGEST by up to 3.5 times. We executed only one process of our algorithm in this experiment. If we had executed multiple parallel processes of our algorithm, we could have achieved higher performance improvement.

\begin{figure}[t]
\centering
\includegraphics[width=3.0in]{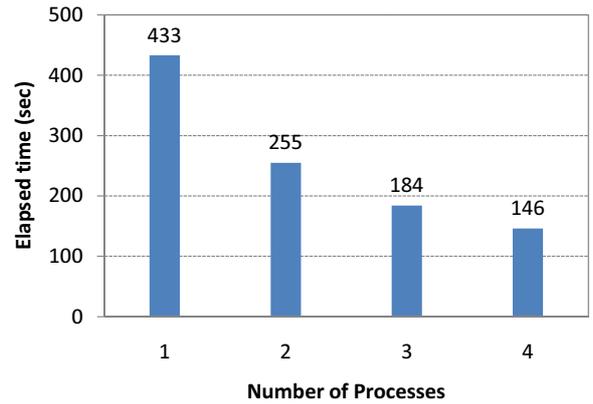}\\
{\footnotesize (a) Using VDB data set.}\\[0.1in]
\includegraphics[width=3.0in]{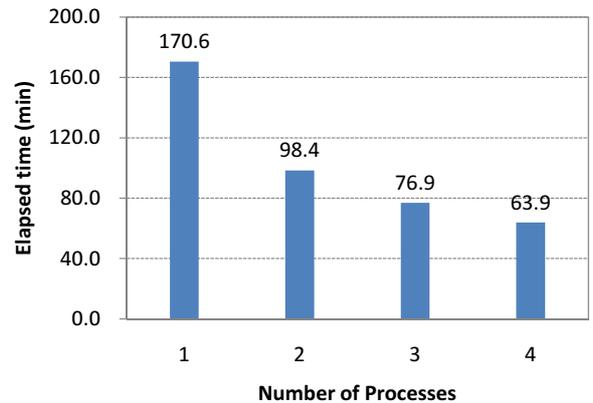}\\
{\footnotesize (b) Using HG18 data set.}
\caption{Result of second experiment: we could obtain performance improvement by up to 3.0 times by running four parallel processes.}
\label{exp2}
\end{figure}

In the second experiment, we ran our algorithm on both VDB and HG18 data sets and compared the elapsed time for various numbers of parallel processes of our algorithm. Figure~\ref{exp2} shows the experimental result. Since the hardware platform has a four-core CPU, we increased the number of parallel processes up to four. Actually, we could have almost no performance improvement by running more than four parallel processes on the same platform. Note that the units of vertical axes are seconds and minutes in Figures~\ref{exp2}(a) and \ref{exp2}(b), respectively. As shown in the figures, we obtained performance improvement by up to 3.0 times by running four parallel processes compared with a single process. We could not obtain four times performance improvement mostly due to inter-process communication and synchronization. Since our algorithm is designed to minimize the effect of disk accesses, it has high potential of more performance improvement by using the advanced CPUs with more cores and faster clock speeds.

\begin{figure}
\centering
\includegraphics[width=3.0in]{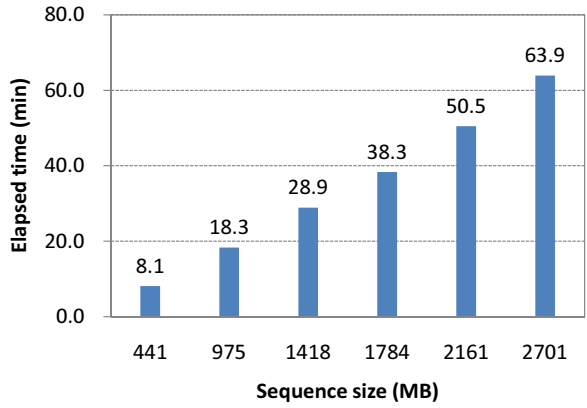}
\caption{Result of third experiment: elapsed time has the linear correlation with the size of genome sequences.}
\label{exp3}
\end{figure}

In the third experiment, we measured the elapsed time of our algorithm for various sizes of genome sequences. We ran four processes on the genome sequences consisting of the first 2, 5, 8, 11, 15, and 24 chromosomes in the human genome sequence. Figure~\ref{exp3} shows the result. As the result of regression analysis on the experimental result, we could find that the elapsed time is almost linearly correlated with the size of genome sequences.

\end{document}